\documentclass[runningheads]{llncs}
\usepackage{graphicx}
\usepackage{lineno}
\usepackage{float}
\usepackage[ruled]{algorithm2e}
\usepackage{times}
\usepackage{helvet}
\usepackage{courier}
\usepackage{graphicx}
\usepackage{amsmath}
\usepackage{multirow}
\usepackage{booktabs}
\usepackage{xcolor}
\usepackage{bm}
\usepackage{mathrsfs}
\usepackage{amssymb}
\usepackage{subfigure}
\usepackage{enumerate}
\newcommand{\squishlist}{
   \begin{list}{$\bullet$}
    { \setlength{\itemsep}{0pt}      \setlength{\parsep}{2pt}
      \setlength{\topsep}{1pt}       \setlength{\partopsep}{0pt}
      \setlength{\leftmargin}{1em} \setlength{\labelwidth}{1em}
      \setlength{\labelsep}{0.5em} } }

\newcommand{\squishend}{
    \end{list}  }

\newcommand{\topcaption}{%
\setlength{\abovecaptionskip}{1pt}%
\setlength{\belowcaptionskip}{1pt}%
\caption}
\begin{document}
\title{Modeling Heterogeneous Edges to Represent Networks with Graph Auto-Encoder}
\newcommand*\samethanks[1][\value{footnote}]{\footnotemark[#1]}
\author{Lu Wang\thanks{equal contribution}\inst{1,2,3} \and
Yu Song\samethanks \inst{1,2,3} \and
Hong Huang\thanks{corresponding author}\inst{1,2,3} \and
Fanghua Ye\inst{4} \and
Xuanhua Shi\inst{1,2,3} \and
Hai Jin\inst{1,2,3}}
% % %
\authorrunning{Lu Wang et al.}
% % First names are abbreviated in the running head.
% % If there are more than two authors, 'et al.' is used.
% %
\institute{National Engineering Research Center for Big Data  Technology and System \and Service Computing Technology and Systems Laboratory \and School of Computer Science and Technology, Huazhong University of  Science  and  Technology, Wuhan, China \and
School of Computer Science, University College London, London, United Kingdom \\
\email{wluluo@gmail.com, yusonghust@hust.edu.cn,  honghuang@hust.edu.cn, smartyfh@outlook.com, xhshi@hust.edu.cn, hjin@hust.edu.cn}}

% \email{lncs@springer.com}\\
% \url{http://www.springer.com/gp/computer-science/lncs} \and
% ABC Institute, Rupert-Karls-University Heidelberg, Heidelberg, Germany\\
% \email{\{abc,lncs\}@uni-heidelberg.de}}
% %
\maketitle              % typeset the header of the contribution
% %
\begin{abstract}
In the real world, networks often contain multiple relationships among nodes, manifested as the heterogeneity of the edges in the networks. We convert the heterogeneous networks into multiple views by using each view to describe a specific type of relationship between nodes, so that we can leverage the collaboration of multiple views to learn the representation of networks with heterogeneous edges.
Given this, we propose a \emph{regularized graph auto-encoders} (RGAE) model, committed to utilizing abundant information in multiple views to learn robust network representations. More specifically, RGAE designs shared and private graph auto-encoders as main components to capture high-order nonlinear structure information of the networks. Besides, two loss functions serve as regularization to extract consistent and unique information, respectively. Concrete experimental results on realistic datasets indicate that our model outperforms state-of-the-art baselines in practical applications.

\end{abstract}

\section{Introduction}

The research of network analysis has made rapid progress in recent years. In fact, network data are usually complex and therefore hard to process. To mine network data, one fundamental task is to learn a low-dimensional representation for each node, such that network properties are preserved in the vector space. As a result, various downstream applications, such as link prediction~\cite{perozzi2014deepwalk}, classification~\cite{wang2016linked}, and community detection\cite{he2015detecting}, can be directly conducted in such vector space. As for learning representations for networks, there are two main challenges that have not yet been fully resolved: 

\noindent \textbf{(1) Preservation of heterogeneous relationships between nodes.} There usually exist diverse and different types of relationships between nodes, leading to the heterogeneity of edges. For example, in the twitter network, four types of relationships may be observed in the interactions between two users, that is one user may retweet, reply, like, and mention another user's tweet. Thus it is reasonable to build four types of edges between the two users with each type of edge corresponding to one type of relationship. Although these edges reflect the similarity between the two users, we can not ignore the slight difference at the “semantic” level. Therefore, taking heterogeneity of edges into consideration for representing such networks is quite significant. In literature, several heterogeneous network embedding approaches~(e.g. PTE~\cite{tang2015pte}, Metapath2vec~\cite{dong2017metapath2vec}, and HIN2Vec\cite{fu2017hin2vec}) have been proposed to represent heterogeneous nodes or edges into the same semantic vector space. However, these methods only learn a final representation for all relationships jointly but ignore the different semantic meanings of edges. Therefore, in order to explore the heterogeneity of edges, it is necessary to learn a relation-specific representation for each type of relationship.         

\noindent \textbf{(2) Preservation of high-order node proximities.}  As described in LINE~\cite{Tang2015LINE}, it defines two loss functions to preserve both 1-st and 2-nd order proximities together. However, it is also meaningful to further integrate the information of k-th-order neighbors for enhancing the representation of nodes with small degrees. Moreover, most existing network embedding methods are equivalent to implicit matrix factorization~\cite{qiu2018network}, which is a shallow model that fails to capture high-order nonlinear proximities between nodes. GraRep~\cite{cao2015grarep} aims to capture the k-th-order proximity by factorizing the k-step~(k=1,2,$\cdots$,K) transition matrices. However, the matrix factorization technique is usually time inefficient and hard to learn nonlinear relationships between nodes. SDNE~\cite{wang2016structural} designs a deep auto-encoder framework to extract the nonlinear structural information of networks, but it still only considers 1-st and 2-nd order proximities without preserving even higher order proximities between nodes. Consequently, to preserve the complex network information, a better solution should leverage high-order nonlinear structural information to yield more robust network representations.

Recently, it has witnessed that multi-view learning is applied successfully in a wide variety of applications, especially for mining heterogeneous data, such as clustering~\cite{kumar2011co}, computer vision~\cite{li2002statistical}, and information retrieval~\cite{pan2014click}. In this regard, we convert heterogeneous edges into multiple views for a network, and solving a multi-view learning problem to learn representations for such networks. To be more specific, we abstract each relationship as a view of the network, reflecting a type of proximity between nodes, thus the original network can be further interpreted as a multi-view network. Finally, we formalize the task as a multi-view network embedding problem. Existing multi-view network embedding methods, such as MVE~\cite{qu2017attention} and MINEs~\cite{ma2018multi}, first learn a single-view network representation using skip-gram model then fuse them directly. Since their fusion strategies, i.e. averaging and adding, are both linear functions, they fail to capture the complex nonlinear information, leading to a sub-optimal result. Besides, there are some works~\cite{shi2018mvn2vec,xu2019multi,zhang2018scalable} learning a unified representation and a view-specific representation for each view simultaneously, but they are shallow models without considering the high-order proximities between nodes.

\begin{figure}[t]
    \centering
    \subfigure[]{
    \includegraphics[scale=0.3]{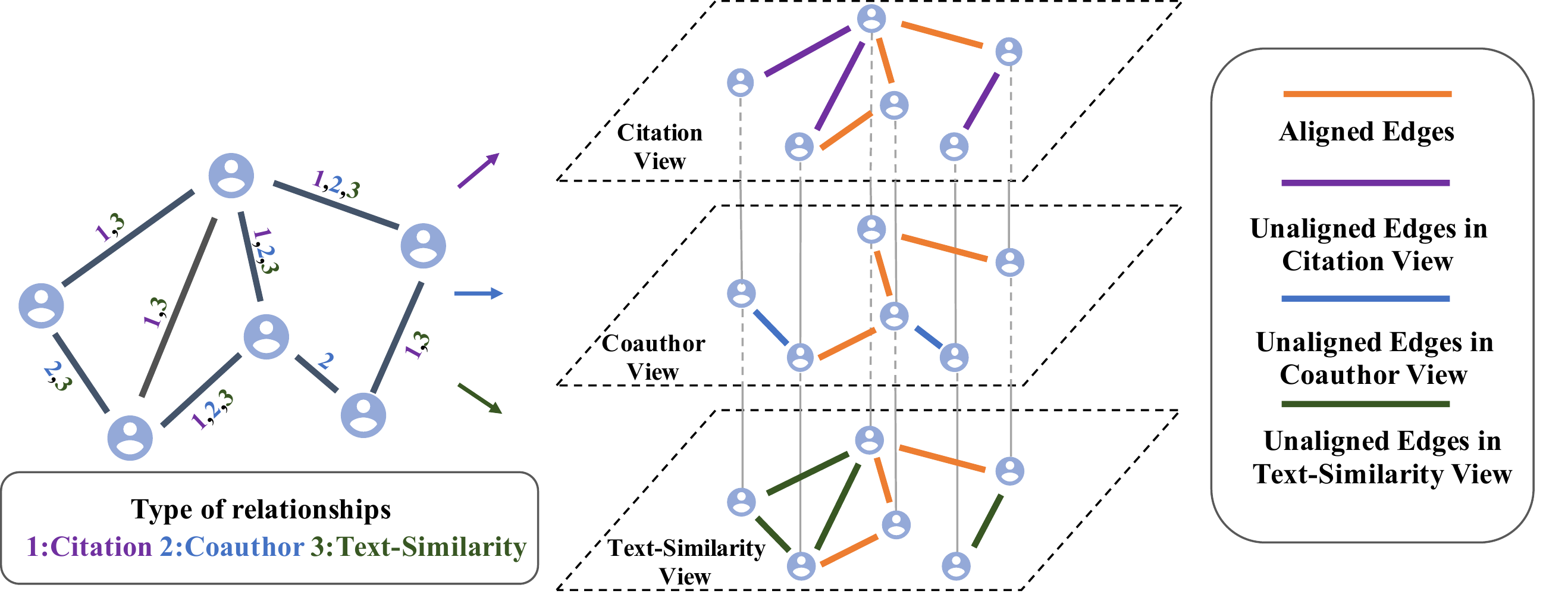}
    }\label{fig1a}
    \quad
    \subfigure[]{
    \includegraphics[scale=0.1]{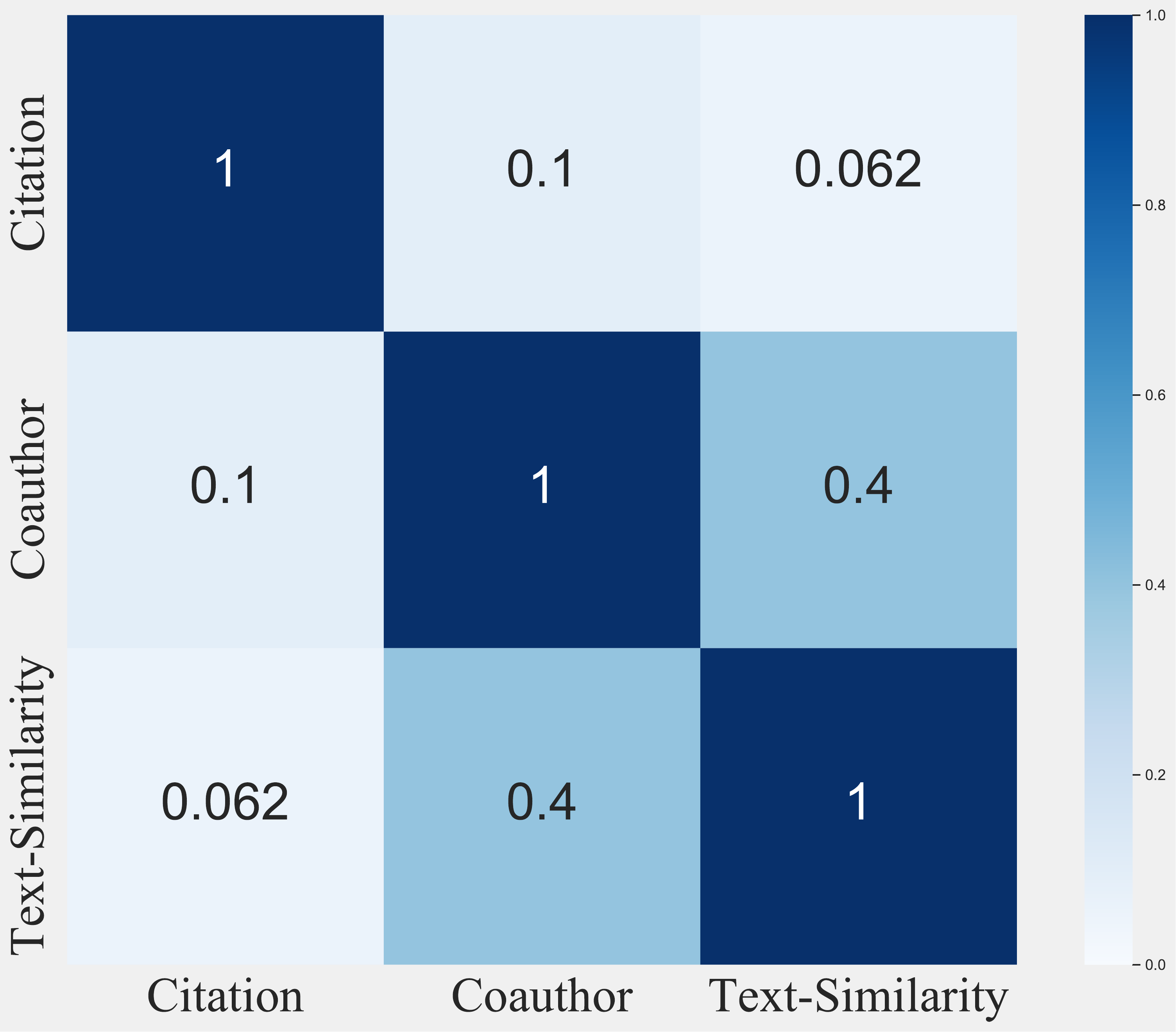}
    }\label{fig1b}
    \vspace{-0.2cm}
    \caption{(a) Illustration of converting heterogeneous relationships to multiple views of the network. (b) Consistent and unique information carried by each pair of AMiner network views.}
    \label{fig1}
    \vspace{-0.5cm}
\end{figure}

Targeting at modeling the heterogeneity of edges and preservation of high-order node proximities for learning network representations, we propose a novel \textbf{R}egularized \textbf{G}raph \textbf{A}uto-\textbf{E}ncoders framework, namely RGAE. To better illustrate our motivation, we first introduce a case study on a multi-view AMiner network~(see details in sec~\ref{exp_set}). As shown in Fig.~\ref{fig1} (a),  it contains two types of information, consistent information and unique information, as its edges are partial aligned as well as partial distinct between different views. Different views may share some consistent information. At the same time, each of them also carries some unique information that others do not have. We further follow a similar method~\cite{shi2018mvn2vec} to perform a statistical analysis. Given a pair of views, the edge sets are $\mathcal{E}_1$ and $\mathcal{E}_2$. We treat the Jaccard coefficient between the two sets as the proportion of consistent information. As we can see in Fig.~\ref{fig1} (b), there exists noticeable consistent information between coauthor and text similarity views while other pairs of views are quite negligible. Thus we conclude that it is unreasonable to preserve only consistent or unique information for multi-view network embedding. As a result, RGAE model aims to preserve consistent and unique information simultaneously, as well as capturing high-order nonlinear proximities between nodes. The contributions of our model are threefold: 

\noindent \textbf{(1)}. In consideration of preserving heterogeneous information of edges as much as possible, we design two kinds of graph auto-encoders to deal with consistent and unique information respectively: one is the shared across view and the other is private to each view. Through these deep and nonlinear graph auto-encoders, our RGAE model is able to represent complex high-order structural information. 

\noindent \textbf{(2)}. We further introduce two regularized loss functions, i.e. the similarity loss and the difference loss, to explicitly avoid the information redundancy of the two types of graph auto-encoders. The similarity loss is used to extract consistent information from shared graph auto-encoders. The difference loss aims to encourage the independence between shared and private graph auto-encoders, so the unique information can also be well preserved at the same time. 

\noindent \textbf{(3)}. To evaluate the performance of the RGAE model, we conduct abundant experiments on four real-world datasets. The experimental results demonstrate that the proposed model is superior to existing state-of-the-art baseline approaches as well as examining the novelty of our model.

\begin{table}[htbp]
% \footnotesize
\centering
\renewcommand\arraystretch{1.0}
\topcaption{Summary of symbols}\label{tab1}
\begin{tabular}{c|c|c|c}
\hline
\textbf{Symbol}                                    & \textbf{Definition}                    & \textbf{Symbol}                                         & \textbf{Definition} \\ \hline
$\mathcal{U}$                             & node set                      & $\mathcal{E}_{i}$                              & edge set of view $i$ \\ \hline
$|V|$                                     & number of views               & $D$                                            & dimension of Y \\ \hline
$N$                                       & number of nodes               & $d$                                            & = ${\lfloor D/(|V|+1)\rfloor}$ \\ \hline
$\textbf{A}_i\in \mathbb{R}^{N \times N}$ & adjacency matrix of view $i$  & $\alpha, \beta, \gamma$                        & hyper-parameters \\ \hline
${\textbf{Y}_{i,p}}\in \mathbb{R}^d$      & private embedding of view $i$ & $\textbf{I}_N \in \mathbb{R}^{N \times N}$         & an identity matrix \\ \hline
${\textbf{Y}_{i,s}}\in \mathbb{R}^d$      & shared embedding of view $i$  & $\tilde{\textbf{A}}_i$                         & = $\textbf{A}_i + I_N$ \\ \hline
${\textbf{Y}_{con}}\in \mathbb{R}^d$      & consistent embedding          & $\tilde{\textbf{D}}_{i}(m,m)$                  & = $\sum_{n} \tilde{\textbf{A}}_{i}(m,n)$  \\ \hline
$\textbf{Y}\in \mathbb{R}^{D}$            & final network embedding       & $\textbf{X}_1 \oplus \textbf{X}_2$             & concatenation in the last dimension   \\ \hline
\end{tabular}
\vspace{-0.5cm}
\end{table}

\section{Problem Formulation and Notations}
We first briefly define a multi-view network, multi-view network embedding and list the main notations used throughout this paper in Table~\ref{tab1}:
\begin{definition}\textbf{Multi-View Network}
A multi-view network is a network defined as $G = \left\{\mathcal{U}, \mathcal{E}_{1}, \mathcal{E}_{2}, \cdots, \mathcal{E}_{|V|} \right\}$, where $\mathcal{U}$ is a node set shared by all views, and $\mathcal{E}_{i}$ $\left ( 1\leq i\leq |V| \right )$ is the edge set of the $i$-th view, which reflects a specific type of relationship between nodes.
\end{definition}

% Obviously, the information contained in multiple views constitutes the complete structure information of the network, so taking multiple views into consideration helps us to learn more robust network representations. Under the circumstance, a formal definition of multi-view network embedding problem is given as follows:

\begin{problem}\textbf{Multi-View Network Embedding}
\qquad Given a multi-view network $G =$ $ \left\{ \mathcal{U}, \mathcal{E}_{1}, \mathcal{E}_{2}, \cdots, \mathcal{E}_{|V|} \right\}$, the multi-view network embedding problem aims to learn a low-dimensional embedding representation $\textbf{Y} \in \mathbb{R}^D$~($D \ll N$). More specifically, an intermediate view-specific embedding representation $\textbf{Y}_{i,p} \in \mathbb{R}^{d}$ is learned to preserve the unique information of view $i$ and a shared embedding representation $\textbf{Y}_{con} \in \mathbb{R}^{d}$ is learned to preserve the consistent information among all views. The final embedding representation $\textbf{Y}$ is obtained from all view-specific embedding representations and the shared embedding representation by an aggregation function. 
\end{problem}

\section{Method}

\begin{figure}[htbp]
\centering
\vspace{-1cm}
\includegraphics[scale=0.3]{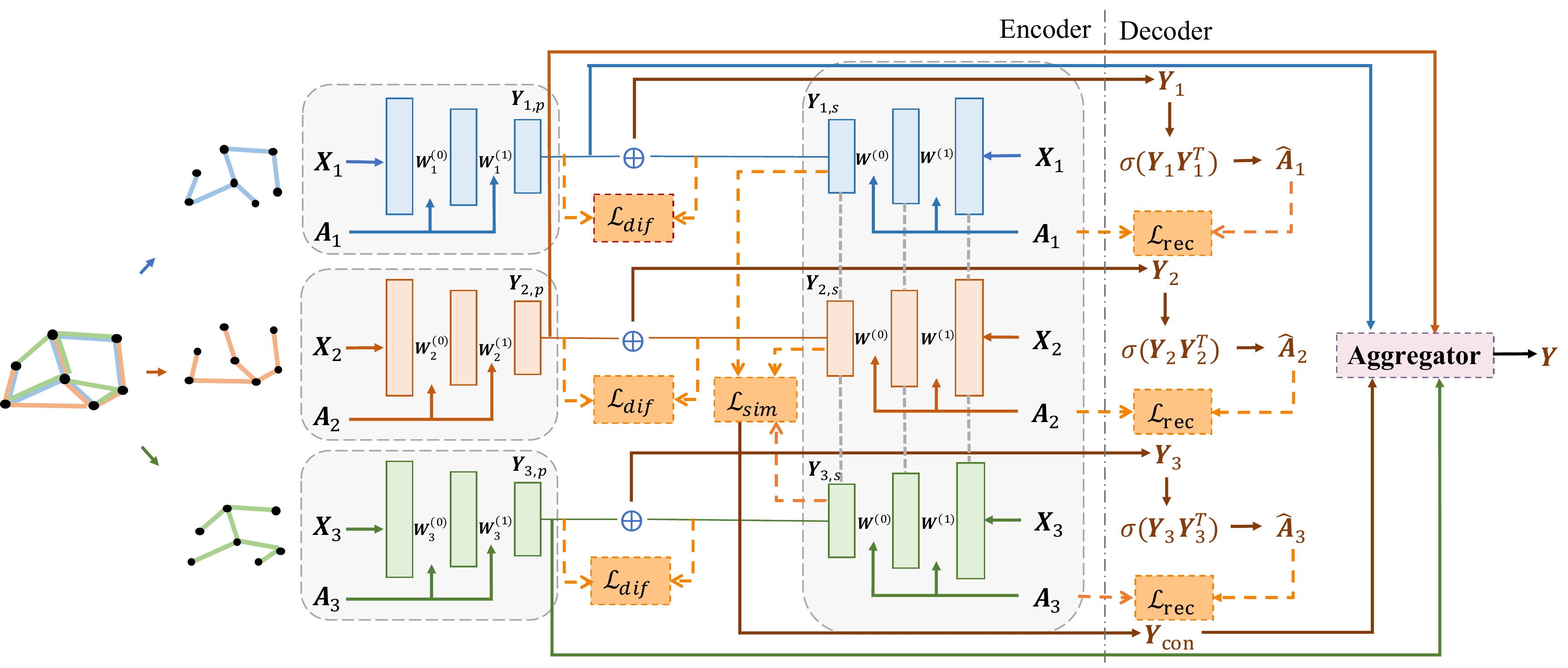}
\vspace{-0.2cm}
\caption{The framework of RGAE. The illustration takes a network with three views as an example.} 
\label{model}
\vspace{-0.5cm}
\end{figure}

In this section, we introduce our proposed \textbf{R}egularized \textbf{G}raph \textbf{A}uto-\textbf{E}ncoders framework, namely RGAE, for tackling the multi-view network embedding problem in detail. An illustrative example of the RGAE model is shown in Fig.~\ref{model}.    

% Overall, the RGAE model contains two types of graph auto-encoders for different purposes. The shared graph auto-encoder allows different views to collaborate together, aiming to extract consistent information among all views. Apart from that, we further utilize a private graph auto-encoder for each view to preserve its unique information. To extract both the consistent information and the unique information, we also add two loss functions, i.e. the similarity loss and the difference loss, to explicitly encourage independence of the two types of graph auto-encoders.
% Finally, the aggregation function is adopted to yield the final robust network representation for downstream tasks. An illustrative example of the RGAE model is shown in Fig.~\ref{model}.    

\subsection{The Shared and Private Graph Auto-Encoders}
\emph{Graph convolutional network}~(GCN)~\cite{kipf2016semi} is built on the idea of message passing, and convolves the representation of the central node with the representations of its neighbors to derive an updated representation of the central node. Our shared and private graph auto-encoders are both motivated as an extension of existing GCN that is able to learn valuable information for graphs. By stacking multiple GCN layers as an encoder and a simple inner production operation as a decoder, the graph auto-encoders in the RGAE model is capable of extracting consistent and unique information in a multi-view network. Specifically, given a multi-view network denoted as $G = \left\{\mathcal{U}, \mathcal{E}_{1}, \mathcal{E}_{2}, \cdots, \mathcal{E}_{|V|} \right\}$, for a specific view $i$, the propagation rule of $l$-th layer in the private encoder is formulated as:
\begin{equation}
    \textbf{Y}_{i,p}^{(l+1)} = \sigma(\tilde{\textbf{D}_i}^{-\frac{1}{2}}\tilde{\textbf{A}}_i\tilde{\textbf{D}}_i^{-\frac{1}{2}}\textbf{Y}_{i,p}^{(l)}\textbf{W}_i^{(l)})\label{eq1}
\end{equation}
where the $\sigma(\cdot)$ is the non-linear activation function. In this paper, we choose $relu$ as activation function in all cases. $\textbf{W}_i^{(l)}$ is the weight matrix, and $\textbf{Y}_{i,p}^{(0)} = \textbf{X}_i$ is the feature matrix for view $i$. Specially, if the node features are not available the $\textbf{X}_i$ will be an identity matrix, as described in \cite{kipf2016semi}. 

% ~\footnote{Note the node set is shared across all views. The shared encoder can also be applied to such multi-view networks that have a few nodes unobserved in each view by padding those unobserved nodes' entries as zero in the adjacency matrix of each view.}

The key point of the shared encoder is that the weight matrices in all layers are shared across different views~\footnote{Note the node set is shared across all views.}, which is clearly different from the private graph auto-encoder. In detail, the propagation rule of the $l$-th layer in the shared graph encoder is formulated as:
\begin{equation}
    \textbf{Y}_{i,s}^{(l+1)} = \sigma(\tilde{\textbf{D}_i}^{-\frac{1}{2}}\tilde{\textbf{A}}_i\tilde{\textbf{D}}_i^{-\frac{1}{2}}\textbf{Y}_{i,s}^{(l)}\textbf{W}^{(l)})\label{eq2}
\end{equation}
Note that the weight matrix $\textbf{W}^{(l)}$ is only shared in view-wise rather than layer-wise. Through this shared architecture we can project all views into the same semantic space so that the process of extracting the consistent information is more interpretable. We can also allow different views to influence mutually and collaborate implicitly. 

The GCN layer is motivated by a first-order approximation of the localized spectral filters on graph-structured data~\cite{defferrard2016convolutional}. In this regard, it is possible to stack multiple GCN layers in both shared encoders and private encoders to capture the high-order proximity between nodes. The final outputs of these stacked shared encoders and private encoders are denoted as $\textbf{Y}_{i,s}$ and $\textbf{Y}_{i,p}$ for each view respectively. During the forward pass, the graph decoder in view $i$ aims to calculate the reconstructed adjacency matrix $\hat{\bf{A}}_i$. In order to utilize the complete information to make a better reconstruction, we first concatenate the outputs of the shared encoder and private encoder for view $i$, then we utilize the inner production operation to yield the reconstructed adjacency matrix, as described in \cite{kipf2016variational}, which is computed as follow:
\begin{equation}
    \textbf{Y}_i = \textbf{Y}_{i,s} \oplus \textbf{Y}_{i,p}, \quad \hat{\textbf{A}}_i = sigmoid(\textbf{Y}_i {\textbf{Y}_i}^\mathrm{T})\label{eq3}
\end{equation}

Since the adjacency matrix preserves the topology information of the graph, it is momentous to minimize the reconstruction loss. It has been demonstrated that minimizing the reconstruction loss is helpful to preserve the similarity between nodes~\cite{salakhutdinov2009semantic}. Due to the sparsity of networks, there exist a great deal of zero elements and the number of zero elements and non-zero elements is extremely unbalanced in the adjacency matrix. As a result, we minimize the reconstruction error by optimizing the Balanced Cross-Entropy loss, which allows the model to pay more attention to the non-zero elements thus ignores the redundant noises from zero elements. For the view $i$, we compute the reconstruction loss as follows:
\begin{equation}
    \mathcal{L}_i^{rec} = \sum_{\tiny{a_i^{(m,n)} \in \textbf{A}_{i}, \hat{a}_{i}^{(m,n)} \in \hat{\textbf{A}}_{i}}} [- a_{i}^{(m,n)} log (\hat{a}_{i}^{(m,n)}) \varsigma - (1-a_{i}^{(m,n)})log(1-\hat{a}_{i}^{(m,n)})]
\label{eq4}
\end{equation}
where the $\varsigma$ is a weighting factor to balance the importance of the non-zero elements, defined as $\frac{\#zero \quad elements}{\#non-zero \quad elements}$ in $\textbf{A}_{i}$.

\subsection{Regularization}
\subsubsection{Similarity Loss}
Intuitively, the consistent information can be extracted from the outputs of the shared encoders. Since we have projected all these outputs into the same semantic space, it is meaningful to make them collaborate to vote for the consistent representation. In this process, we encourage the consistent representation $\textbf{Y}_{con}$ to be similar to the shared representation $\textbf{Y}_{i,s}$ of each view as much as possible. As the importance of views may be different, we further allow the model to assign different weights to them. Taking all these into consideration, we introduce the following similarity loss to regularize the extraction process: 

\begin{equation}
    \mathcal{L}^{sim} = \sum_{i=1}^{|V|} \lambda_{i}^{\gamma} \Vert \textbf{Y}_{con} - \textbf{Y}_{i,s} \Vert_F^2, \quad \sum_{i=1}^{|V|} \lambda_{i} = 1, \lambda_{i} \geq 0 \label{eq5}
\end{equation}
where $\lambda_{i}$ is the weight for view $i$, and $\gamma$ moderates the weight distribution. By learning proper weights, the extraction process can let the consistent representation focus on the most informative views. Naturally, the consistent representation is calculated as the weighted combinations of the outputs of the shared encoders, which illustrates the collaboration between different views.   
\subsubsection{Difference Loss}
In order to preserve the unique information, the difference loss is also introduced to encourage the  isolation between consistent embeddings and unique embeddings. As the consistent information and unique information have essential differences, they should be distinguished clearly to avoid the information redundancy. In other words, the shared embeddings and private embeddings should describe the information of multiple views in different perspectives, thus we define the difference loss via an orthogonality constraint between the private embedding and shared embedding in each view:
\begin{equation}
    \mathcal{L}_i^{dif} = \Vert \textbf{Y}_{i,s} \odot \textbf{Y}_{i,p} \Vert_F^2 ,  \quad i = 1,2,\cdots,|V| \label{eq6}
\end{equation}
where the $\odot$ is the row-wise inner production. Obviously, the difference loss will drive the shared embeddings to be orthogonal with the private embeddings, thus they will be as dissimilar as possible. In this way, the shared and private encoders are able to encode different aspects of the multi-view network. In this paper, we treat the output of the private graph encoder for each view as its private representation. 

\subsection{The Aggregation Process}
As introduced above, our RGAE model includes the three types of losses, i.e. the reconstruction loss, the similarity loss, and the difference loss. In order to train these losses jointly, the overall loss of our proposed model is summarized as follow:
\begin{equation}
    \mathcal{L} = \sum_{i=1}^{|V|} \mathcal{L}_i^{rec} + \alpha * \mathcal{L}^{sim} + \beta * \sum_{i=1}^{|V|} \mathcal{L}_i^{dif}  
    \label{eq7}
\end{equation}
where $\alpha$ and $\beta$ are hyper-parameters to control the importance of similarity loss and difference loss respectively. Up to now, we have obtained the representations of consistent and unique information. Finally, we design an aggregation process to yield the final network representation, which can be illustrated as:
\begin{equation}
    \textbf{Y} = Aggregator(\textbf{Y}_{con},\textbf{Y}_{1,p},\cdots,\textbf{Y}_{|V|,p})\label{eq8}
\end{equation}
The aggregator should be able to integrate both the consistent and unique information effectively, and it can be add, average, pooling and some other designed functions. In this paper, we choose concatenation as the aggregation function since it has been proven to be useful and efficient in many existing network embedding methods~\cite{li2018multi,shi2018mvn2vec,Tang2015LINE}. As shown in Table~\ref{tab1}, the total dimension $D$ has been assigned to each graph auto-encoder equally, thus after the concatenation process the final network embedding will still satisfy $\textbf{Y}\in \mathbb{R}^{D}$.

\subsection{Implementation}
In practice, we utilize Tensorflow for an efficient GPU-based implementation of the RGAE model.  Then the parameters of RGAE model except $\lambda_i$ can be efficiently optimized automatically with back propagation algorithm. To save space, we omit details here. Since the sparsity of network data, we use sparse-dense matrix multiplication for Eqs.~\eqref{eq1} and \eqref{eq2}, as described in ~\cite{kipf2016semi}. Specially, for the view weight $\lambda_i$ in Eq.~\eqref{eq5}, we follow the same method~\cite{cai2013multi} to update it. Let's denote $\Vert \textbf{Y}_{con} - \textbf{Y}_{i,s} \Vert_F^2$ as $\bm{B}_i$, then Eq.~\eqref{eq5} is equivalent to $\sum_{i=1}^{|V|} \lambda_{i}^{\gamma}\bf{B}_i - \xi (\sum_{i=1}^{|V|} \lambda_{i}-1)$, where $\xi$ is Lagrange multiplier. By taking the derivative of this formula with respect to $\lambda_i$ as zero, we can obtain the update rule of $\lambda_i$: $\lambda_i \leftarrow \frac{(\gamma\bf{B}_i)^{\frac{1}{1-\gamma}}}{\sum_{i=1}^{|V|}(\gamma\bf{B}_i)^{\frac{1}{1-\gamma}}}$.  It is efficient to use one parameter $\gamma$ for controlling the distribution of view weights during the optimization process dynamically. According to the update rule, we would assign equal weights to all views when $\gamma$ closes to $\infty$. When $\gamma$ closes to 1, the weight for the view whose $\bm{B}_i$ value is smallest will be assigned as 1, while others are almost ignored since their weights are close to 0. 
The pseudo code is shown in Algorithm.~\ref{algorithm}.

% Briefly speaking, the computational complexity of RGAE model is linear to the number of edges in different views, the hidden size, the number of views and the size of total embedding dimensions. 

\begin{algorithm}[htbp]
\LinesNumbered
\KwIn{$G = \left \{\mathcal{U}, \mathcal{E}_{1}, \mathcal{E}_{2}, ..., \mathcal{E}_{V} \right \}$,$D$,$K$,$\alpha$,$\beta$,$\gamma$ and hidden size for each encoder layer;
}
\KwOut{The network representation $\bm{Y}$ for the network G.}
\For{i=1,2,$\cdots$,$|V|$}{
According to Eq.~\eqref{eq1}, construct private graph encoder for view $i$;\\
According to Eq.~\eqref{eq2}, construct shared graph encoder for view $i$;\\
According to Eqs.~\eqref{eq3} and \eqref{eq4}, construct graph decoder and calculate $\mathcal{L}_i^{rec}$;\\
}
According to Eq.~\eqref{eq5} and Eq.~\eqref{eq6}, calculate $\mathcal{L}^{sim}$ and $\mathcal{L}_i^{dif}$;\\
\Repeat
{\text{Convergence}}
{
Update parameters in RGAE model via optimizing the $\mathcal{L}$ in Eq.~\eqref{eq7};\\
}
According to Eq.~\eqref{eq8}, obtain the final network representation $\bm{Y}$;\\
\Return Network representation $\bm{Y}$.
\caption{The pseudo-code of RGAE Model}
\label{algorithm}
\end{algorithm}

\section{Experiments}

\subsection{Experimental Setup} \label{exp_set}
We select four multi-view network datasets in different fields. The statistic analysis is shown in Table~\ref{tab2}.

\begin{table}[ht]
\centering
% \footnotesize
\renewcommand\arraystretch{1.0}
\setlength{\tabcolsep}{1.0mm}
\topcaption{Overview of datasets}\label{tab2}
\begin{tabular}{c|cccccc}
\hline
Task                             & Dataset & Views & Nodes & Edges   & Labels & Type       \\ \hline \hline
\multirow{2}{*}{Multi-class Node Classification} & AMiner  & 3     & 8,438  & 2,433,356 & 8      & Academic   \\
                                     & PPI     & 6     & 4,328  & 1,661,756 & 50     & Biological \\ \hline
\multirow{1}{*}{Multi-label Node Classification} & Flickr  & 2     & 34,881 & 3,290,030 & 171     & Social     \\ \hline
\multirow{1}{*}{Link Prediction}     & YouTube & 4     & 5,108  & 3,263,045 & -      & Social     \\\hline
\end{tabular}
\vspace{-0.5cm}
\end{table}

% \squishlist
% \item AAA
% \item BBB
% \squishend

\squishlist
	\item AMiner~\cite{tang2008arnetminer}: AMiner network is an academic network representing the relationships between authors. It consists of three views: author-citation, co-authorship, and text similarity. Text similarity between two authors is calculated by TF-IDF from titles and abstracts in their papers. An author establishes connections with his top ten similar authors and we only preserve authors in eight research fields as \cite{dong2017metapath2vec}. The research fields are treated as node labels.
	
	\item PPI~\cite{franceschini2012string}: The PPI network is a human protein-protein interaction network. Six views are constructed based on the co-expression, co-occurrence, database, experiment, fusion, and neighborhood information. Gene groups are treated as node labels.
	
	\item Flickr~\cite{tang2009relational}: It is a social network of online users on Flickr with two views. One view is the friendship network among bloggers. The other is a tag-proximity network in which a node connects with its top 10 similar nodes according to their tags. We treat community memberships as node labels.
	
	\item YouTube~\cite{yang2015defining}: It is a social network consists of four views: the friendship, the number of common friends, the number of common subscribers, and the number of common favorite videos between two users.
\squishend

In order to evaluate the effectiveness of RGAE, we compare our model with three types of baselines. The \textbf{single-view} based baselines include:

% \begin{itemize}
\squishlist
    \item Deepwalk~\cite{perozzi2014deepwalk}: It is a well-known baseline for network embedding. We set the number and the length for each node as 80 and 40 respectively following the recommendations of the original paper. The window-size is set as 10.
    \item GraRep~\cite{cao2015grarep}: It aims to capture the k-order proximities by factorizing the k-step transition matrices. We set k as 5.
    \item SDNE~\cite{wang2016structural}: It utilizes the auto-encoders to preserve the neighbor structure of nodes. The first-order and second-order proximity are proposed to preserve the global and the local network structure. We set the number of layers as 3, and the hidden size as [800,400,128]. 
    \item GAE~\cite{kipf2016variational}: It stacks GCN layers as an encoder and the inner production operation as a decoder. The reconstruction loss helps it to capture structural information in an unsupervised manner. We set the number of layers and hidden sizes same as SDNE.
\squishend
The \textbf{heterogeneous} network embedding methods include:
\squishlist
\item  PTE~\cite{tang2015pte}: It is a heterogeneous network embedding method which can also be used to jointly train the embedding, because multi-view network is a special type of heterogeneous network. We set the number of negative samples as 5.
\item Metapath2vec~\cite{dong2017metapath2vec}: It utilizes meta-paths guided random walk to generate the node sequences then uses skip-gram model to learn the node representations.  We set the number, the length of walks and window size same as deepwalk. We perform experiment using one of all possible meta-paths at a time, and report the best result.
\squishend

The \textbf{multi-view} based baselines include:
\squishlist
    \item Deepwalk-con: It applies Deepwalk to get a $d$ dimensional representation for each view then concatenates these representations from all $K$ views to generate a unified representation with $K \ast d$ dimensions.
    \item MultiNMF~\cite{liu2013multi}: It is a multi-view matrix factorization algorithm, which extracts consistent information by a joint matrix factorization process. 
    \item MVE~\cite{qu2017attention}: It combines single view embeddings by weights learned from attention mechanism to construct a multi-view network embedding. We set the parameters of random walk and skip-gram model same as Deepwalk, and other parameters are same as the original paper.
    \item MNE~\cite{zhang2018scalable}: It combines the information of multiple view by preserving a high dimensional common embedding and a lower dimensional embedding for each view. The dimensions of the additional vectors are set as 10.
    \item MTNE-C~\cite{xu2019multi}: It combines the common embedding and node-specific embedding of each node to be a complete embedding for the closeness measurement. We follow the default parameter setting in the original paper.
\squishend

For RGAE and all baselines except Deepwalk-con, the embedding dimension is set as 128. The number of graph auto-encoder layers is set as 3, and two hidden layers' dimensions are set as 800 and 400 respectively. Both $\alpha$ and $\beta$ are selected from [0.1,0.3,0.5,0.7,1.0,1.5,2.0], and $\gamma$ is selected from [0.05,0.5,5,10,50,100,500]. The learning rate is selected from [0.001,0.01,0.1]. As node features are not available for our datasets the feature matrix will be an identity matrix. We treat the node embedding learned by various methods as feature to train linear classifiers for multiclass classification, and train one-vs-rest classifiers for multilabel classification. For link prediction, we use the cosine similarity between node pairs as features to train a logistic classifier to predict the link existence. Follow the setting in \cite{qu2017attention}, we use other three views to train embeddings and predict the link existence in friend view. To generate negative edges, we randomly sample an equal number of node pairs which have no edge connecting them.  We report the best results among multiple views for single-view based baselines. To guarantee a fair comparison, we repeat each method ten times and the average metrics are reported.

\begin{table}[!hbt]
\centering
% \footnotesize
\renewcommand\arraystretch{1.0}
\setlength{\tabcolsep}{1.0mm}
\topcaption{ Node classification results w.r.t. Micro-F1(\%) and Macro-F1(\%) with different training ratio. '-' means out of memory error.}\label{tab3}
\begin{tabular}{c|c|c|cc|cc|cc}
\hline
\multirow{2}{*}{Datesets} & \multirow{2}{*}{Category}      & \multirow{2}{*}{Methods} & \multicolumn{2}{c|}{0.1} & \multicolumn{2}{c|}{0.3} & \multicolumn{2}{c}{0.5} \\ \cline{4-9} 
                          &                                &                          & Micro       & Macro      & Micro       & Macro      & Micro      & Macro      \\ \hline \hline
\multirow{12}{*}{AMiner}  & \multirow{4}{*}{Single-View}   & Deepwalk                 & 69.9       & 68.4      & 74.3       & 73.3      & 75.1      & 74.3      \\
                          &                                & GraRep                   & 23.3       & 20.5      & 44.8       & 41.8      & 61.9      & 60.6      \\
                          &                                & SDNE                     & 64.8       & 62.5      & 70.3       & 68.0      & 70.8      & 69.4      \\
                          &                                & GAE                      & 60.4       & 54.8      & 62.5       & 57.7      & 63.6      & 59.3      \\ \cline{2-9} 
                          & \multirow{2}{*}{Heterogeneous} & PTE                      & 52.9       & 46.9      & 56.6       & 52.7      & 58.1      & 55.2      \\
                          &                                & Metapath2Vec             & 70.6       & 70.1      & 75.3       & 73.5      & 76.2      & 74.9         \\ \cline{2-9} 
                          & \multirow{6}{*}{Multi-View}    & Deepwalk-con             & 61.4       & 59.0      & 74.2       & 72.6      & 76.1      & 74.9      \\
                          &                                & MultiNMF                 & 57.4       & 52.6      & 66.4       & 64.1      & 66.8      & 62.8      \\
                          &                                & MVE                      & 73.6       & 72.7      & 78.8       & 77.5      & 78.9      & 77.6      \\
                          &                                & MNE                      & 73.6       & 72.2      & 79.2       & 77.8      & 79.6      & 78.1      \\
                          &                                & MTNE-C                   & 54.5       & 48.8      & 57.2       & 53.9      & 58.6      & 55.2      \\ \cline{3-9} 
                          &                                & RGAE                    & \textbf{74.9}& \textbf{73.3}      & \textbf{80.6}       & \textbf{79.7}      & \textbf{82.0}      & \textbf{80.9}      \\ \hline \hline
\multirow{12}{*}{PPI}     & \multirow{4}{*}{Single-View}   & Deepwalk                 & 8.9       & 4.2      & 10.9       & 6.1      & 12.1      & 7.3      \\
                          &                                & GraRep                   & 4.0       & 2.0      & 5.1       & 3.1      & 13.1      & 10.0      \\
                          &                                & SDNE                     & 11.8       & 10.7      & 14.7       & 13.4      & 17.6      & 15.0      \\
                          &                                & GAE                      & 9.5       & 4.3      & 12.3       & 8.0      & 13.7      & 9.1      \\ \cline{2-9} 
                          & \multirow{2}{*}{Heterogeneous} & PTE                      & 12.8       & 9.5      & 19.7       & 11.7      & 22.0      & 14.0      \\
                          &                                & Metapath2Vec             & 13.4       & 10.0     & 20.2      & 12.8       & 22.3      & 15.7         \\ \cline{2-9} 
                          & \multirow{6}{*}{Multi-View}    & Deepwalk-con             & 9.9       & 6.2      & 11.9       & 8.5      & 13.6      & 9.9      \\
                          &                                & MultiNMF                 & 15.3       & 11.9      & 17.8       & 15.2      & 20.3      & 17.5      \\
                          &                                & MVE                      & 11.7       & 9.9      & 12.1       & 10.6      & 13.3      & 10.8      \\
                          &                                & MNE                      & 13.3       & 11.8      & 14.1       & 12.2      & 15.6      & 12.1      \\
                          &                                & MTNE-C                   & 3.4       & 1.6      & 4.0       & 2.0      & 6.2      & 3.5      \\ \cline{3-9} 
                          &                                & RGAE                    & \textbf{19.0}       & \textbf{15.1}      & \textbf{24.4}       & \textbf{21.0}      & \textbf{25.0}      & \textbf{21.3}      \\ \hline \hline
\multirow{12}{*}{Flickr}  & \multirow{4}{*}{Single-View}   & Deepwalk                 & 51.7       & 32.1      & 51.9       & 27.6      & 53.2      & 27.8      \\
                          &                                & GraRep                   & 52.4       & 32.2      & 53.8      & \textbf{35.0}      & 55.9      & 35.8      \\
                          &                                & SDNE                     & 47.6       & 32.1      & 48.2       & 32.6      & 49.6      & 30.5      \\
                          &                                & GAE                      & 34.5       & 9.1      & 37.0       & 10.4      & 38.4      & 11.1      \\ \cline{2-9} 
                          & \multirow{2}{*}{Heterogeneous} & PTE                      & 55.7       & 30.4      & 56.4       & 34.3      & 56.2      & 31.0      \\
                          &                                & Metapath2Vec             & 55.7       & 30.8      & 56.6       & 33.9         & 56.7         & 32.2         \\ \cline{2-9} 
                          & \multirow{6}{*}{Multi-View}    & Deepwalk-con             & 51.9       & 32.6      & 52.5       & 28.2      & 53.7      & 28.3      \\
                          &                                & MultiNMF                 & -           & -          & -           & -          & -          & -          \\
                          &                                & MVE                      & 52.0       & 32.5      & 53.0       & 28.9      & 54.3      & 28.8      \\
                          &                                & MNE                      & 52.4       & \textbf{33.1}      & 53.5       & 29.9      & 54.8      & 29.8      \\
                          &                                & MTNE-C                   & 23.9       & 5.2      & 23.3       & 4.8      & 22.9      & 4.6      \\ \cline{3-9} 
                          &                                & RGAE                    & \textbf{56.7}       & 32.9      & \textbf{57.6}       & 33.7      & \textbf{58.4}      & \textbf{36.2}      \\ \hline
\end{tabular}
\vspace{-0.5cm}
\end{table}

\subsection{Experimental Results}
\subsubsection{Node Classification}
We evaluate the performance of our method and three categories of baselines using the Micro-F1 and Macro-F1 scores. Table~\ref{tab3} shows the comparison on three datasets. As can be seen, our RGAE model outperforms all baselines except for Macro-F1 on Flickr dataset. For example, on AMiner dataset, it achieves a sustainable performance gain of 1\%, 2\%, and 3\% with the percentage of training data increasing. It is noted that RGAE always outperforms GAE consistently, which shows that with making good use of information from multiple views, we are indeed able to learn a robust representation for a multi-view network. The superiority of RGAE over SDNE further verifies that it is reasonable to model the heterogeneity of edges. Although GraRep captures high order proximities between nodes, the matrix factorization process makes it hard to preserve non-linear network information, which is not compared with our model. One may see that the existing multi-view network embedding approaches are also not comparable to our RGAE model. The reason is that either they are not possible to consider the uniqueness of each view, like Metapath2Vec and MVE, or they are not possible to capture high-order proximities between nodes, such as MTNE-C and MNE. All these observed results show that the RGAE model can indeed capture more complete non-linear information from multiple views.

\begin{table}[!hbt]
\centering
\renewcommand\arraystretch{1.0}
\setlength{\tabcolsep}{1.5mm}
\topcaption{Link Prediction results on YouTube dataset w.r.t. ROC\_AUC Score(\%) and Average Precision Score (AP)(\%) with different training ratio}\label{tab4}
\begin{tabular}{c|c|cc|cc|cc}
\hline
\multirow{2}{*}{Category}      & \multirow{2}{*}{Methods} & \multicolumn{2}{c|}{0.1} & \multicolumn{2}{c|}{0.3} & \multicolumn{2}{c}{0.5} \\ \cline{3-8} 
                               &                          & ROC\_AUC     & AP        & ROC\_AUC     & AP        & ROC\_AUC     & AP       \\ \hline \hline
\multirow{4}{*}{Single-View}   & Deepwalk                 & 74.4        & 73.6     & 74.7        & 74.0     & 78.4        & 77.2    \\
                               & GraRep                   & 80.2        & 79.6     & 80.3        & 79.8     & 80.7        & 80.0    \\
                               & SDNE                     & 81.8        & 82.7     & 82.3        & 83.0     & 85.0        & 85.3    \\
                               & GAE                      & 77.0        & 77.7     & 77.3        & 78.2     & 80.3        & 79.6    \\ \hline
\multirow{2}{*}{Heterogeneous} & PTE                      & 69.5        & 63.8     & 70.1        & 64.8     & 69.1        & 64.7    \\
                               & Metapath2Vec             & 78.5        & 73.8     & 80.6        & 75.8     & 81.9        & 79.7    \\ \hline
\multirow{6}{*}{Multi-View}    & Deepwalk-con             & 78.9        & 78.0     & 79.8        & 78.9     & 84.7        & 83.1    \\
                               & MultiNMF                 & 80.3        & 80.2     & 81.9        & 82.3     & 82.2        & 82.8    \\
                               & MVE                      & 82.0        & 82.4     & 83.0        & 82.8     & 83.4        & 83.1    \\
                               & MNE                      & 82.3        & 82.7     & 83.3        & 83.5     & 84.1        & 84.6    \\
                               & MTNE-C                   & 52.4        & 53.0     & 62.3        & 62.9     & 66.1        & 65.8    \\ \cline{2-8} 
                               & RGAE                    & \textbf{82.7} & \textbf{83.2} & \textbf{85.5} & \textbf{85.2} & \textbf{86.3} & \textbf{85.9}    \\ \hline
\end{tabular}
\vspace{-0.5cm}
\end{table}

\subsubsection{Link Prediction}
We select the YouTube dataset to verify the performance of link prediction. Table~\ref{tab4} shows that the RGAE model significantly outperforms all baseline methods. The results verify again that RGAE indeed can preserve the abundant information in multi-view networks. It is noticeable that the SDNE even outperforms all multi-view and heterogeneous network embedding approaches. By designing two kinds of graph auto-encoders, RGAE utilizes both consistent and unique information from multiple views to describe the node proximity in a detailed way, which achieves better performance than SDNE. As a result, we conclude that RGAE is able to explore the structural properties of multi-view networks.

\begin{figure*}[htbp]
    \centering
    \subfigure[Citation]{
    \includegraphics[scale=0.115]{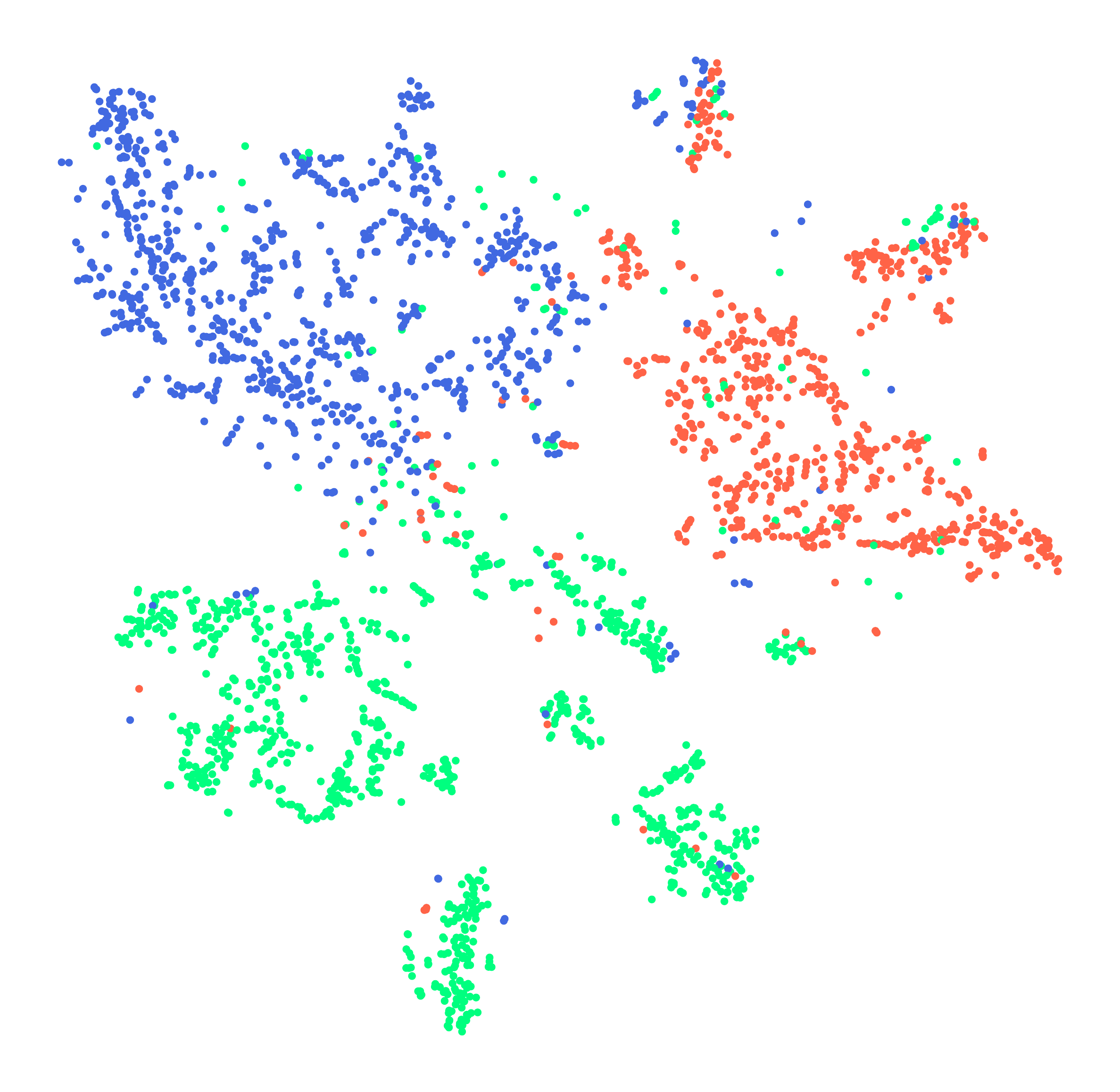}
    }
    \subfigure[Coauthor]{
    \includegraphics[scale=0.115]{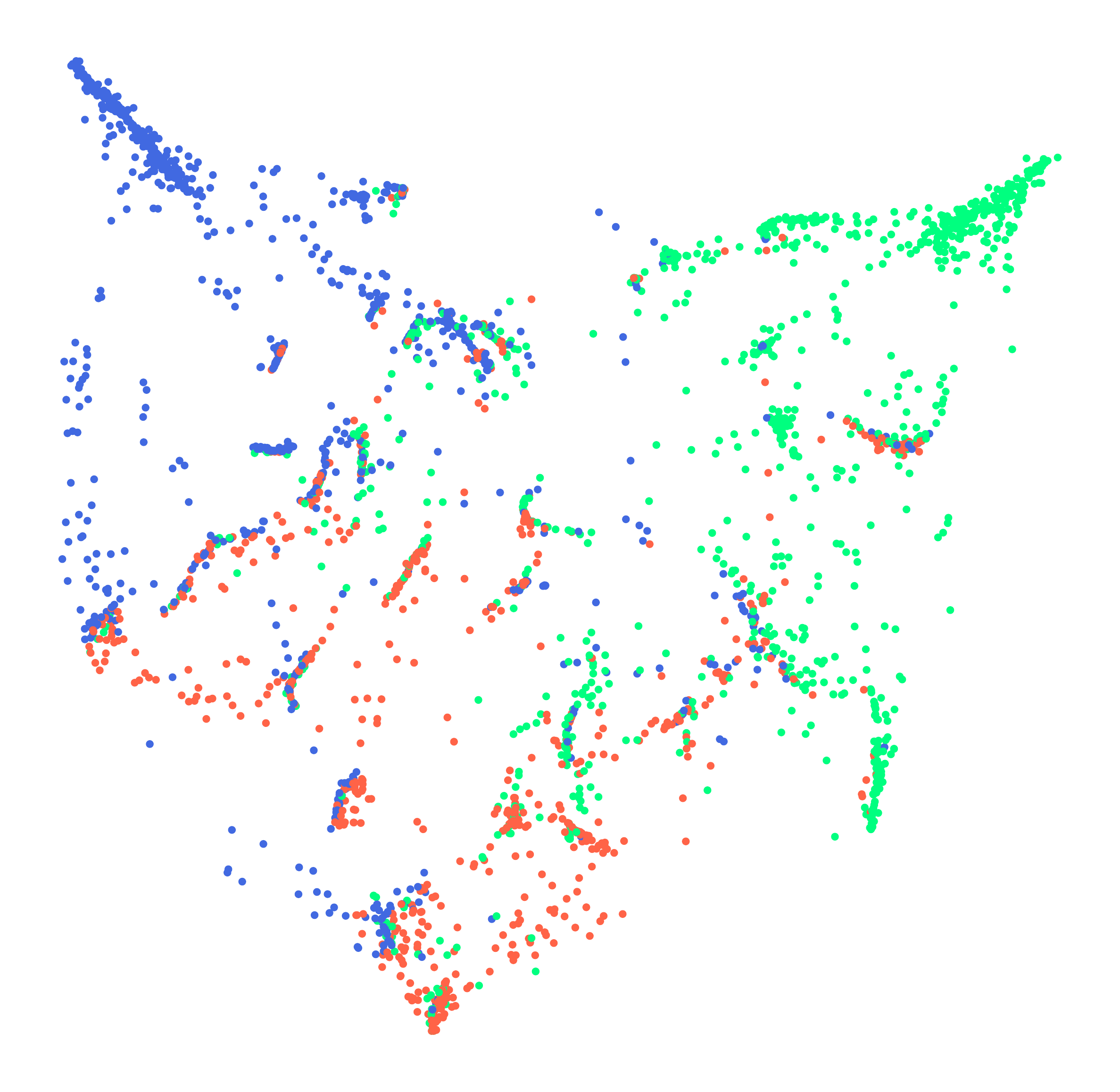}
    }
    \subfigure[Text-Similarity]{
    \includegraphics[scale=0.115]{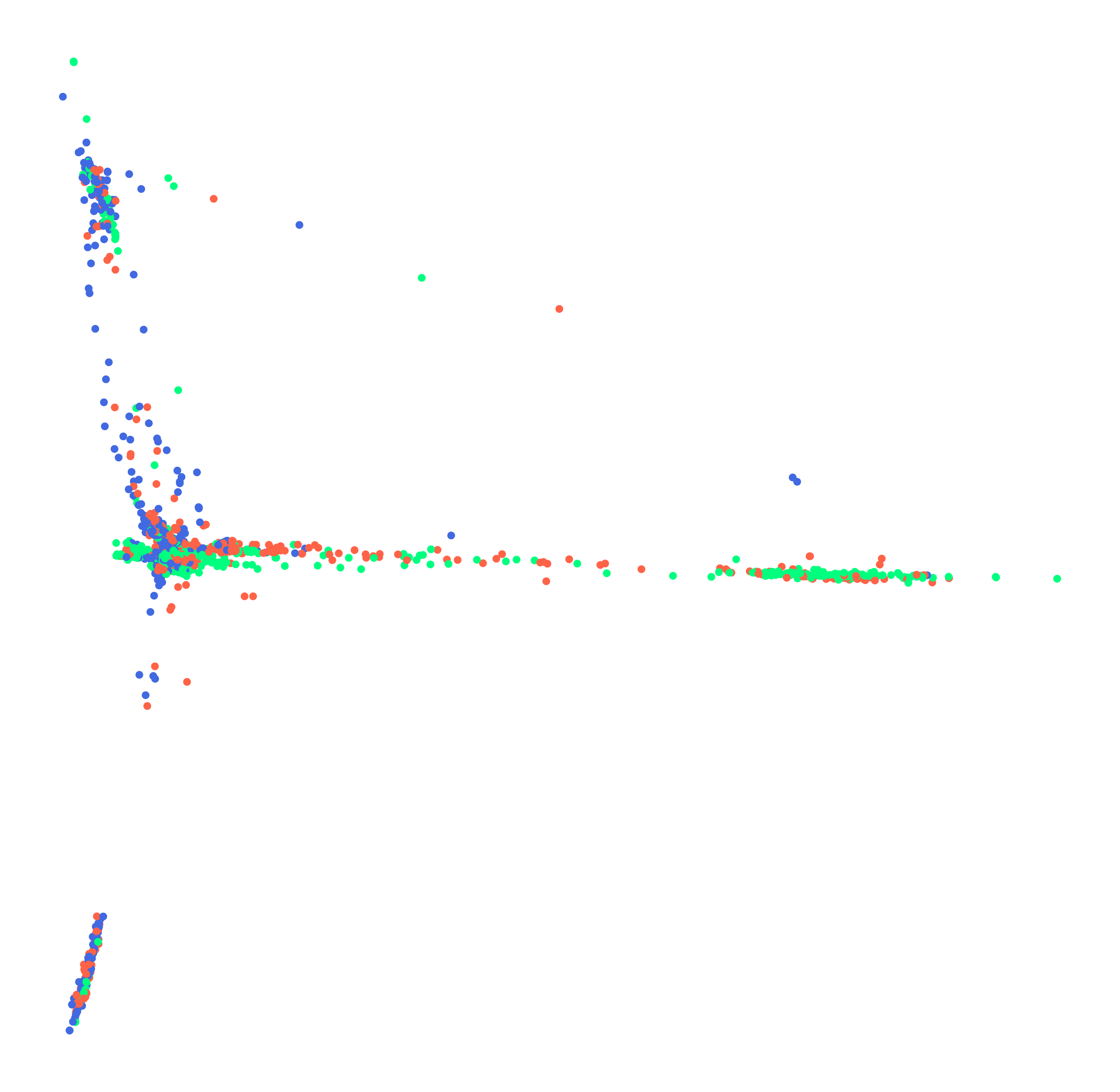}
    }
    \subfigure[RGAE]{
    \includegraphics[scale=0.115]{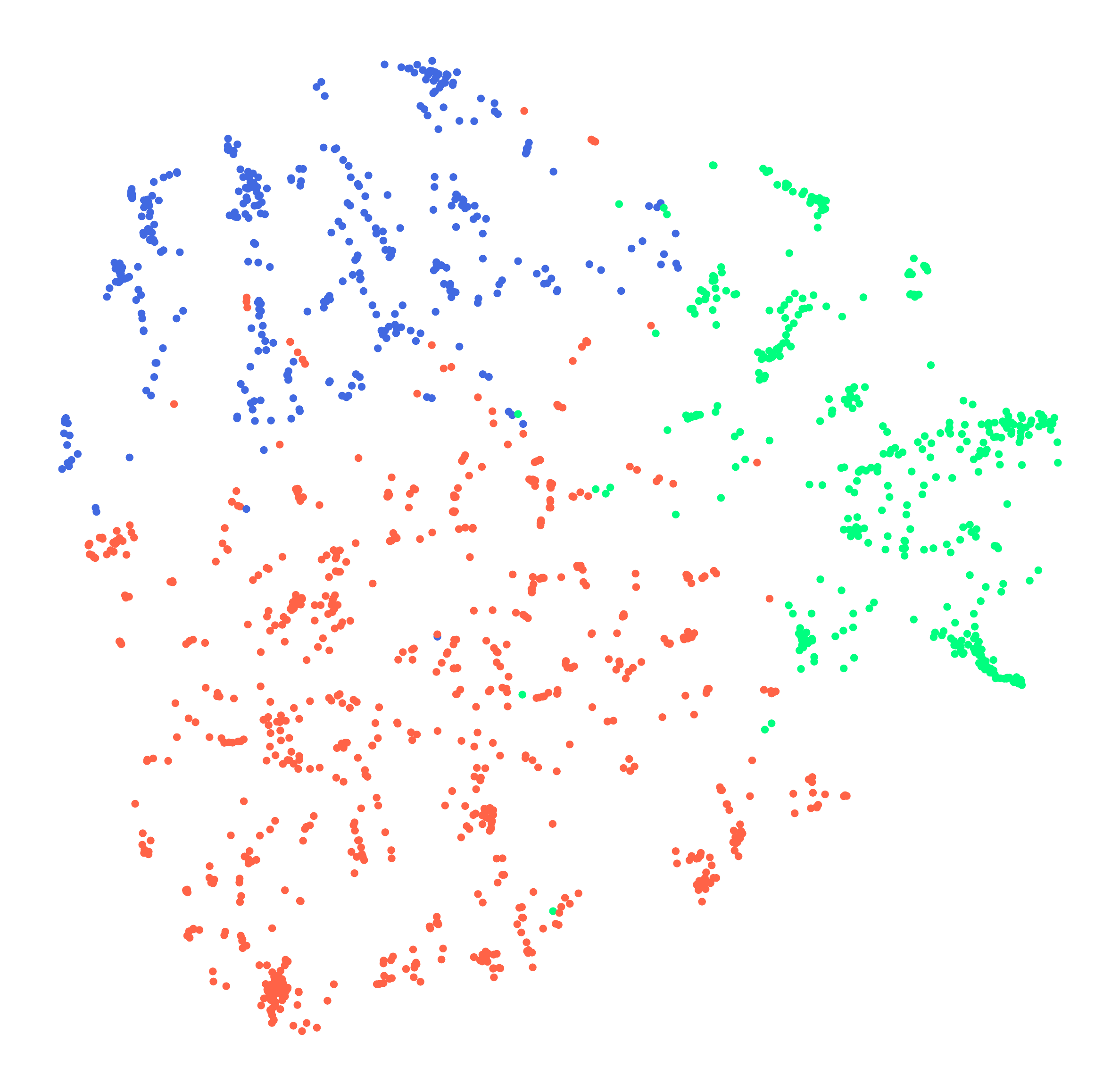}
    }
    \vspace{-0.2cm}
    \caption{2d t-SNE Visualization for AMiner Dataset. Each point represents a node and colors represent labels. red : computational linguistics; blue : computer graphics; green : theoretical computer science}
    \vspace{-1cm}
    \label{vis}
\end{figure*}

\subsubsection{Network Visualization}
We project the embeddings of AMiner dataset onto 2d vectors with t-SNE~\cite{maaten2008visualizing}. Fig.~\ref{vis} shows the network visualizations of the RGAE model as well as each view's visualization obtained by its shared and private encoders. The difference between RGAE model and the single-view model is that single-view model lacks not only the constraints of the loss function to divide the consistent information and unique information in a view, but also the cooperation and supplementary between different views. In order to make the visualization results more clear and legible, we select three from the eight categories of nodes for visualization, and each color represents a research field. We can see that our multi-view based approach works better than learning each view individually. Citation view may achieve relatively good representation effect, but there are still a few nodes that have not been assigned to the correct cluster. Therefore, it still needs more useful information from other views to complement and properly correct it to get a robust representation. The visualization of RGAE, by contrast, separates the three research fields clearly, which illustrates the necessity of modeling heterogeneous edges with consideration of all types of relationships.

\subsection{Influence of Loss Functions}
\begin{figure}[ht]
    \centering
    \subfigure[Node classification on AMiner dataset w.r.t Micro-F1(\%)]{
    \includegraphics[width=4cm,height=3cm]{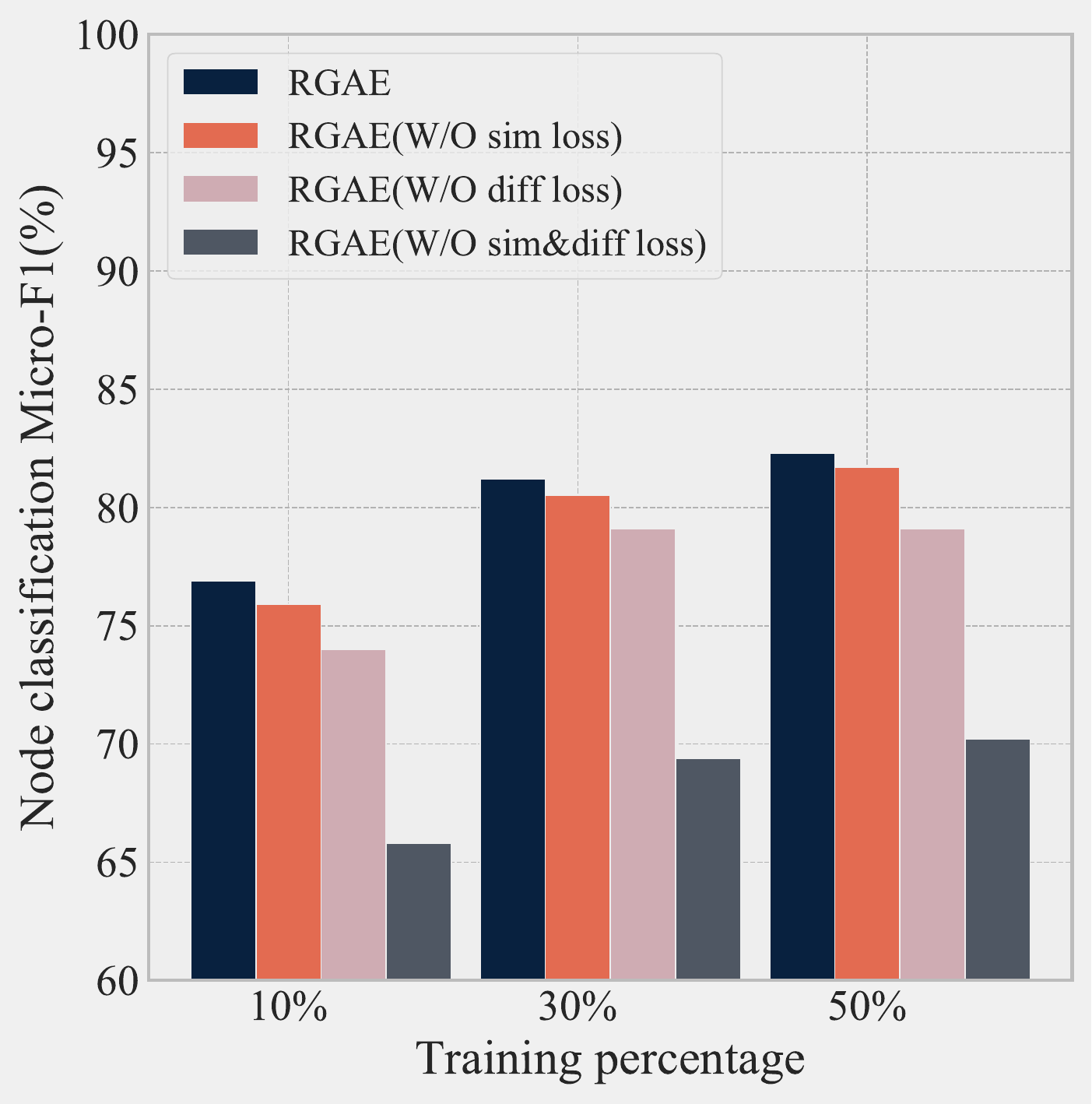}
    }
    \qquad
    \qquad
    \subfigure[Node classification on AMiner dataset w.r.t Macro-F1(\%)]{
    \includegraphics[width=4cm,height=3cm]{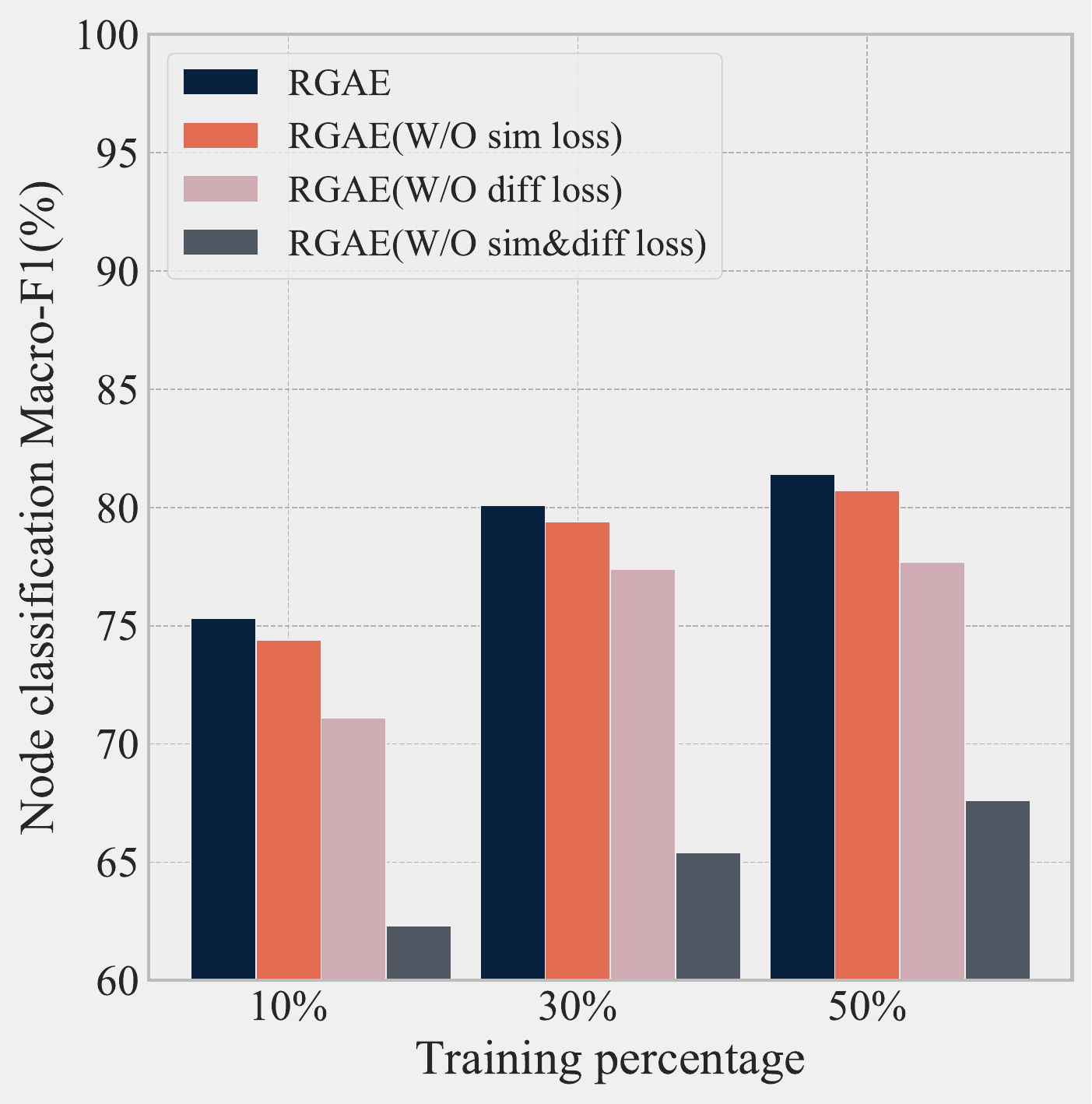}
    }
    \vspace{-0.2cm}
    \caption{The effectiveness of the similarity loss and difference loss in our RGAE model}
    \vspace{-1cm}
    \label{loss}
\end{figure}

The visualization results have proven the importance of both consistent and unique information. In this part, we research the effect of the loss functions. In our RGAE model, there exist two loss functions, i.e. the similarity loss and the difference loss, that regularize the processes of extracting the consistent and unique information respectively. To evaluate the influences of the two loss functions, we remove similarity loss, difference loss, and both of them respectively, and show the performance in Fig.~\ref{loss}. The histogram clearly shows the importance of the two loss functions for our RGAE model. When we remove the similarity loss function, there is slight decline in performance. Because without similarity loss function, the quality of consistent information will be affected. Whereas there is relatively little consistent information among the views, and the proportion of the dimensions of the common representation in the final representation is small, so that the performance declination will not be quite severe. When there is no difference loss, there will be a noticeable decrease in performance, because the isolation between different view's specific information becomes worse without the regularization of difference loss. Moreover, if we remove the similarity loss and difference loss simultaneously, the performance of the RGAE model declines further dramatically. All these observations can demonstrate the necessity of the similarity loss and difference loss, but the degree of influence varies between the two losses.

\subsection{Parameter Sensitivity}
\begin{figure}
    \centering
    \subfigure[Parameter sensitivity on AMiner node classification]{
    \includegraphics[scale=0.18]{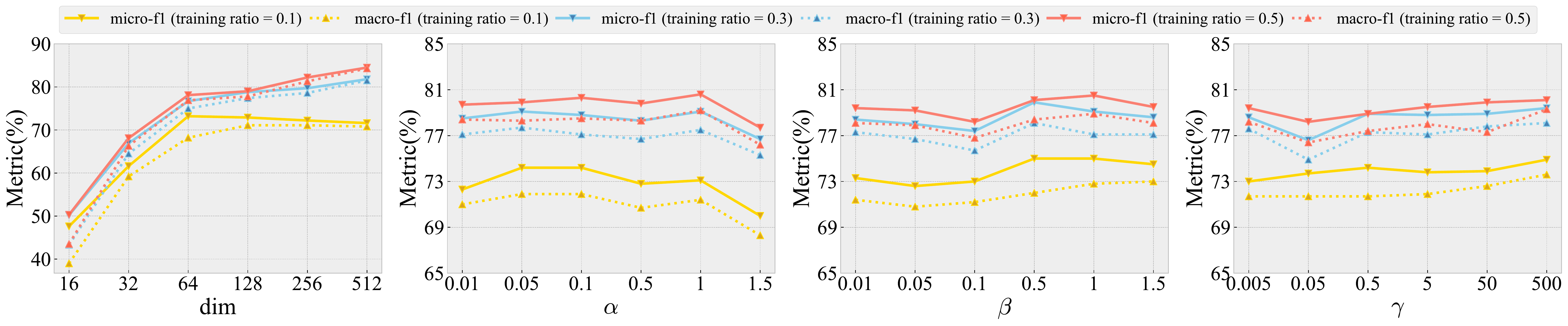}
    }
    \subfigure[Parameter sensitivity on YouTube link prediction]{
    \includegraphics[scale=0.18]{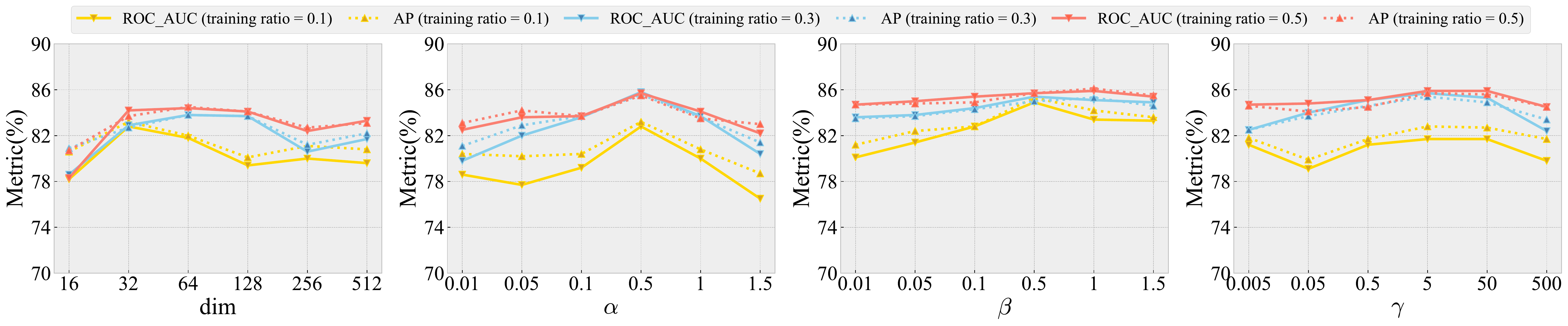}
    }
    \vspace{-2mm}
    \caption{Performance of the RGAE models under varying hyper-parameters}
    \label{para}
    \vspace{-3mm}
\end{figure}

With results presented in Fig.~\ref{para}, we focus on the parameter sensitivity of RGAE model, including the number of embedding dimensions, $\alpha$, $\beta$, and $\gamma$. We perform node classification on AMiner dataset and link prediction on YouTube dataset to evaluate the parameter sensitivity. To explore the contributions of these parameters, we fix others to evaluate the effect of one parameter at a time on the experimental results.

Overall, different datasets and tasks have different sensitivities to embedding dimensions. On AMiner dataset, the performance increases with the dimension increasing then stabilizes when the dimension reaches 64. While on the YouTube dataset, the model performs well when the dimension is 32 and the performance decreases slightly when the dimension continues to increase. Compared with the AMiner dataset, the Youtube dataset achieves good results in lower dimensions. When the proportion of the training data set is small, a large number of dimensions tend to cause overfitting. 

The curves of the experimental metrics with the parameters $\alpha$ or $\beta$ are not monotonic. Their overall trends are both first rising and then falling. Because when the proportion of similarity loss function and difference loss function are too large, the proportion of reconstruction loss will be weakened, which will affect the representation abilities of graph auto-encoders. As for $\gamma$, we find that it actually influences the results for both tasks. As we can see, it is more suitable to set the value of $\gamma$ larger than 5. 
\vspace{-3mm}

\vspace{-8mm}
\section{Related Work}
\subsubsection{Network Embedding: }
Network embedding is dedicated to mapping nodes in a network into a low-dimensional vector space for preserving structural information. Earlier studies such as Deepwalk~\cite{perozzi2014deepwalk}, node2vec~\cite{grover2016node2vec}, and Struc2vec~\cite{ribeiro2017struc2vec} use skip-gram model to preserve network structures through neighborhood sampling. Traditional deep neural networks also get widespread attention because of its nonlinear underlying structure. SDNE~\cite{wang2016structural}, SiNE~\cite{wang2017signed}, and Deepcas~\cite{li2017deepcas} have a strong advantage in retaining the highly nonlinear structure of the network.More recent methods adopt graph neural network to perform convolutional operations on graphs. GCN~\cite{kipf2016semi}, GATs~\cite{velivckovic2017graph}, and GraphSAGE~\cite{hamilton2017inductive} are all representative works as end-to-end approaches for network representation learning. These studies are directed at homogeneous networks. Heterogeneous network embedding has also attracted attention because of its practical significance. PTE~\cite{tang2015pte} is an extension method of LINE on heterogeneous networks.  Besides, Metapath2vec~\cite{dong2017metapath2vec}, HIN2vec~\cite{fu2017hin2vec}, and RHINE~\cite{lu2019auto} use meta path to capture the structure and semantic information in heterogeneous networks.
\vspace{-5mm}
\subsubsection{Multi-view Learning: }
Another related work is about multi-view learning. Some traditional multi-view learning algorithms, such as co-training~\cite{kumar2011co}, co-clustering~\cite{yao2017revisiting}, and cross-domain fusion~\cite{franco2005fusion} analyze multi-view networks for specific tasks. MVE~\cite{qu2017attention}, MINES~\cite{ma2018multi}, MVNE~\cite{sun2018multi}, and mvn2vec~\cite{shi2018mvn2vec} account for the first-order collaboration to align the representations of each node across views.
For these studies, the models responsible for learning the network representation of each view are shallow so that they cannot capture the high-order non-linear network structure. With that in mind, we consider using deep neural networks to replace the shallow models as the basic components to embed the network. ACMVL~\cite{lu2019auto} uses multiple auto-encoders to learn the specific features of each view and map all specific features to the same potential space. But it requires a supervised network to help the auto-encoder optimize its parameters. Compared with it, our model is totally unsupervised to solve the multi-view network embedding problem.

\vspace{-5mm}
\section{Conclusion}
\vspace{-2mm}
In this paper, we explore how to model the heterogeneity of edges by solving a multi-view network embedding problem and propose a novel RGAE model. More specifically, our model makes use of two types of graph auto-encoders to extract consistent and unique information of views respectively, and innovatively proposes two loss functions to distinguish these two types of information. Experimental results not only indicate the superiority of the proposed model but also investigate the contributions of two loss functions. In the future, we plan to apply the framework to more applications. A meaningful direction is to use multi-view learning to represent general heterogeneous networks, that is, the nodes and edges of the network have multiple types at the same time.

\vspace{-3mm}
\section*{Acknowledgement}
\vspace{-0.2cm}
The research was supported by National Natural Science Foundation of China (No. 61802140) and Hubei Provincial Natural Science Foundation (No. 2018CFB200).
\vspace{-0.3cm}
%
% ---- Bibliography ----
%
% BibTeX users should specify bibliography style 'splncs04'.
% References will then be sorted and formatted in the correct style.
%
% \bibliographystyle{splncs04}
% \bibliography{mybibliography}
%

\bibliographystyle{splncs04}
\bibliography{ref.bib}
\end{document}